\documentclass[aps,twocolumn,twoside,floatfix,nofootinbib,a4paper]{revtex4}
\usepackage{graphicx}
\usepackage{marginnote}

\usepackage{hyperref}
\usepackage{enumitem,kantlipsum}
\usepackage[usenames,dvipsnames]{color}
\usepackage{amsmath}
\usepackage{amssymb}
\usepackage{amsfonts}
\usepackage{mathrsfs}
\usepackage{bm}
\usepackage[all]{xy}
\usepackage{hyperref}
\usepackage{tikz-cd}

\newcommand{\beq}{\begin{equation}}
\newcommand{\eeq}{\end{equation}}
\newcommand{\bea}{\begin{eqnarray}}
\newcommand{\eea}{\end{eqnarray}}
	
\newcommand{\ea}{\end{align*}}

\newcommand{\bma}{\begin{pmatrix}}
\newcommand{\ema}{\end{pmatrix}}

\usepackage[percent]{overpic}
\usepackage{overpic}

\begin{document}
\title{Memory recall by controlling chaos}
\author{Fan Zhang} 
\affiliation{Gravitational Wave and Cosmology Laboratory, Department of Astronomy, Beijing Normal University, Beijing 100875, China}
\affiliation{Advanced Institute of Natural Sciences, Beijing Normal University at Zhuhai 519087, China}

\date{\today}

\begin{abstract}
\begin{center}
\begin{minipage}[c]{0.9\textwidth}
By incorporating feedback loops, that engender amplification and damping so that output is not proportional to input, the biological neural networks become highly nonlinear and thus very likely chaotic in nature. Research in control theory reveals that strange attractors can be approximated by collection of cycles, and be collapsed into a more coherent state centered on one of them if we exert control. We speculate that human memories are encoded by such cycles, and can be retrieved once sensory or virtual cues, acting as references, enable feedback controls that nucleates the otherwise chaotic wandering mind.
 \end{minipage} 
 \end{center}
  \end{abstract}
\maketitle

\raggedbottom

\section{Walking on thin ice}
The human brain appears to exhibit some level of flexible chaotic behavior (see Appendix \ref{sec:prevalence} for some general arguments on why this is to be expected), as it is capable of sparks of inspiration, or sudden and seemingly random transfers to related concepts, and not just rigid derivation. It is also noted (see e.g., \cite{grebogi1997impact}) that chaotic dynamics could help form the hierarchy of time scales involved in brain chemistry and psychology, or that it would allow for very quick switch between mental states. Indeed, many studies have now proposed that strange attractors play a vital role in neurological processes, see e.g., \cite{Yao1990ModelOB,doi:10.1073/pnas.91.19.9027,tsuda_2001}.  
Experimentally also, the electroencephalographic studies on cerebral electrovoltage fluctuations have been argued by some \cite{haken1982evolution} to exhibit chaotic features that changes in accordance with brain state. 
In particular, near phase transition between orderly and disorderly states of any system, one often observes critical phenomena that signify long range correlations, and this has been seen in the brain via functional magnetic resonance imaging and magnetoencephalography (see e.g., \cite{criticality}). The weight of the circumstantial evidences thus points to the brain operating at the seam between order and anarchy, somehow maintained in a precarious balance, much like how fighter jets operate at the edge of aerodynamic instability, a feature that affords much agility. 

Such a balancing act is quite possible according to control theory (see Appendix \ref{sec:control} and e.g., \cite{ControlChaos}), which demonstrates that mechanisms are readily available for tweaking a dynamical system in and out of chaos. Indeed, previous studies \cite{doi:10.1073/pnas.91.19.9027} have argued that it is possible to impinge arbitrary generic controls, like any loud noise or flash of light, to drive the brain into some more coherent state, that represents a more attentive or alert state of mind (see also Sec.~16.4 in \cite{scholl2008handbook}, the pacemaker there is the attention attracting ``loud bang''; also experimentally, a reduction in neuron firing rate variability has been seen when a stimulus is applied \cite{churchland2010stimulus,he2013spontaneous,schurger2010reproducibility,xue2010greater,doi:10.1073-pnas.1418730112}). The alert state could perhaps be another strange attractor with a lower Hausdorff dimension than that corresponding to the resting state, but it would unlikely reach all the way down to nearly everywhere differentiable level of regularity, which would trivialize the mental state too much and become pathological, e.g., regular periodic spikes in electroencephalographic signals is seen during episodes of epilepsy \cite{doi:10.1142/S0218127491000531}. 

It is worth noting that this extreme condition can be induced through viewing certain flashing images, so the regularization of the mental strange attractor must be somehow related to how our brain processes external stimuli, and arguably the most important aspect of which is memory recall. Specifically, while a short-circuited brain with too much connectivity \cite{2004PhRvL..93x4103Z} between different clusters of neurons or brain regions could suffer from overwhelming synchronization, so the entire mental state collapses into a genuine periodic motion entrained onto the rhythm of the epilepsy-inducing flashing image; a normal brain may only experience a partial collapse, whereby only the mental state's projection onto some sub-dimensions of the phase space for the neural pathways, corresponding to those regions of the brain involved in memory tasks, collapses to periodic motion that encodes memories. Indeed, some level of periodicity has been experimentally observed to correlate with memory tasks, see e.g., \cite{freeman1987simulation,howard2003gamma,pesaran2002temporal,tallon1998induced,wimmer2016transitions,spitzer2010oscillatory,jokisch2007modulation,lee2005phase,raghavachari2001gating,tesche2000theta,gevins1997high}. Such collapses into periodic cycles is the hallmark of control theory approaches that suppress chaos\footnote{The cycles were originally embedded in the strange attractor but repelling, then stabilized by the control. See Appendix \ref{sec:control}.}, so we propose that beyond taking our mental state into alertness with an arbitrary stimuli\footnote{A phase transition in the mental state perhaps induced by increased energy and oxygen supply, or in other words changes in the environmental parameters, which incidentally enhances Hebbian \cite{hebb2005organization} learning capacity.}, a more elaborate feedback-control mechanism, taking a more detailed and informative stimuli (can be external or internal, such as sentential premises, coming from other regions of the brain) as reference, could be responsible for memory recall.  

\section{Finding needle in haystack}
Specifically, in the fashion of a vastly simplified caricature, the memory of say, a vase that we are beholding, would resemble the reference neural dynamics presently imprinted onto our perceptual neural sub-network\footnote{That is, the reference is also a lower dimensional projected trajectory, but onto those subdimensions associated with perception rather than memory. Nevertheless, the dimensions of the two projections can match so they can be compared.}. But that memory would be hiding as an unstable projective cycle in the strange attractor of our suitably focused state of mind. To find it, we can establish a negative feedback to penalize mental state evolutions that increase the distance between its projected trajectory and the reference system (with the memory and perception dimensions suitably mapped), thereby partially\footnote{\label{fn:cylinder}Note, the originally unstable cycles embedded in the strange attractor only need to be stabilized in those dimensions related to memory tasks; the evolution in the remaining dimensions can stay divergent, so we have a less demanding task for the control mechanism. For intuition, one can imagine a 3-D phase space where the memory is associated with say, the x-y plane, then the projective cycle can be any trajectory wrapped around a cylinder that transversely intersects with that plane into a closed orbit. The motion along the length of the cylinder is unconstrained, and can very well be chaotic once we incorporate higher dimensionality. In fact, even along the transverse x-y directions, divergences are allowed to happen at some segments of the cylinder. It is only the stability after averaging across the entire length of the cylinder that we deem necessary. This further relaxes the criteria for the effectiveness of the control mechanism.} collapsing the strange attractor into the target projective cycle, which is now also stabilized, ready to be accessed repeatedly during further analysis. With this feedback approach, the added controls vanish when the deviation has been vanquished, so we are confident that the resulting memory is in fact an orbit of the original uncontrolled dynamics, even as the nearby phase space regions receive a catastrophic make over. Or in other words, the memory is least doctored by the recall process itself, and can remain faithful. In contrast, non-feedback controls tend to persist and are thus more intrusive.

On the other hand, when the stimulant is not exact or if the memory has deteriorated (e.g., neurons died or are recycled for other memories), the feedback control system becomes ineffective in properly stabilizing any projective cycle\footnote{We have now the situation where $\mathcal{F}_c  \approx \mathcal{F}$ (see Appendix \ref{sec:control}) but not exactly equal. The difference between the two is termed ``modeling error'' in the control theory context. Rigorous proof of the effectiveness of the Model-reference Adaptive control method (cf., Theorem 4.2 in \cite{ControlChaos}) assumes no such error.}. Instead, a few orbits that each partially overlaps (small time-averaged error function) with the reference get incompletely stabilized, and become less strongly repelling (now with less positive Lyapunov exponents) pseudo-projective cycles\footnote{Note here we mean a trajectory that is almost a cycle -- if one extracts the spectrum for such a signal, there will be a prominent peak -- but is nevertheless still building up to eventually depart the region. We do not mean a quasi-periodic orbit, in the sense of possessing incommensurate periods.} (cf., the attractor ruin concept of chaotic itinerancy \cite{tsuda_2001,doi:10.1063/1.3076393}). The mental trajectory traversing their neighborhoods would then exhibit intermittent behaviour with both laminar and chaotic phases, as per \cite{Freeman2004SimulationOC,skarda_freeman_1987}. 
This is a failure for control theory, but a triumph for animal survival, as a new situation can be juxtaposed against more than one similar past experiences, so one can pick the comparatively better course of action. 

Furthermore, sometimes even more novel situations may be encountered for which we had no similar prior experience at all, which really shouldn't admit any memory recall. In such situations, the feedback control cannot even produce pseudo-projective cycles, and the mental state remains chaotic, in accordance with the experimental observations of \cite{skarda_freeman_1987}, that sees reactions to known smells consist of regular time series signals, while those to unknown smells remain erratic. The meandering of the mind in this situation does not need to be the same as when stimuli is absent though. As long as the coupling between the sensory and memory subdimensions are in place, the stimuli will have an impact. Specifically, there can be other simpler, less precise control influences at work. One such example is the periodic parametric perturbations presented in Ref.~\cite{1999PhLA..254..275M}, which shows that orbits can at times be stabilized into limit cycles, by simply periodically perturbing some accessible parameter of the system at frequencies that are low order rational multiples of the natural frequency of the originally unstable orbit. This means if our novel stimulant happens to share some frequency domain structure with a past memory, then that memory could be accidentally evoked, even though it really doesn't match the stimulant in any obvious way. Nevertheless, it sharing frequency domain features might make it a useful analogy for making sense of the new situation. For example, in the next section, we propose devising logical relations out of transfers between projective cycles, for which there are likely characteristic time scales that manifest in frequency domain, so two scenarios taking after each other there may be alike in their e.g, causal structures between constituent events. 

\section{Weaving web of guesses}
Our proposal echoes the associative memory recall mechanism of an Attractor or Chaotic \cite{doi:10.1073-pnas.79.8.2554,AIHARA1990333} Neural Network, in that the memory end up being recalled is a (quasi-)trapping region in the mental dynamics (stabilized projective cycle for us, fixed point for Attractor and strange attractor for Chaotic). However, it is antithetical to those models in the querying mechanism. Essentially, since both the dynamical system and the initial state together determine where the evolution ends up at, there are two knobs one can turn, and we choose differently. Those models use the cue to place the initial condition in the attracting basin of the correct attractor, of a fixed (during recall) dynamical system. In contrast, we leave the initial conditions free to be whatever a person's present state of mind is, but instead alter the dynamical system on the fly, according to the cue and via the feedback controls, to manufacture the correct trapping region in situ. (Note further that while in some refinement proposals, control techniques are applied to Attractor or Chaotic networks, e.g., as in \cite{1997PhRvE..56.6410S,doi:10.1142-S0218127403006819,he2007partial}, their task there were rather minor, to e.g., increase storage capacity or help pin the recall output.)

There are subsequently some salient functional features on which the models diverge. For example, the projective cycles are only stabilized with effort, when controls are maintained, and revert to repelling once the controls are gone. Therefore, in contrast to those genuine attractor based memories that trap the mind forever (if nothing is done to deliberately jot), one's mind is automatically released to handle other tasks. This same tendency to avoid trapping also could aid escape from any local minima (false memories) of some Lyapunov function (should one exists), which is a problem plaguing Attractor or Chaotic neural networks \cite{1987LNP...275..429A}, thereby does away with the need to artificially introduce noises \cite{Annealing,1997PhRvE..56.6410S}. 

Furthermore, one of the reasons why we prefer a projective cycle-based encoding is that since the cycles densely fill out the attractors, there is in principle infinite memory capacity (modulo accidental degeneracies in the projected-out non-memory dimensions of the phase space), seemingly mirroring human long term memory, and in contrast to entire attractor-based encodings that suffer from small capacity issues \cite{NIPS2016_eaae339c}. 
Another reason is that we envisage also a dynamical systems underlying for the relationships between the memory items. As such, pre-stabilization projective cycles embedded within the same attractor are more closely interwoven than different attractors, each of which will trap the mental dynamics indefinitely so there is no intrinsic transiting between them\footnote{Note the pre-stabilization projective cycles cannot be sub-attractors, since the strange attractors are by definition minimal.}. 

The way human brains establish relationships between different memory items likely underlies the logical operations between mathematical concepts \cite{Houd2003NeuralFO}, so an elucidation of this brain function is most useful for scientists, as it paves the way for replicating, in an automated artificial research assistant, human's ability to craft theories. It is not very difficult to concoct relations between unstable projective cycles (encoding concepts) that could plausibly underpin logical relations. For example, with causation $\Rightarrow$, we could have two pre-stabilization projective cycles connected up by a generous flux of transiting paths (cf., the transfer between the two wings on a Lorenz attractor for intuition) that are unidirectional, so one always precedes the other, but never the opposite way around. Once a memory recall event happens at the departure projective-cycle, it becomes mostly stable, but with fluctuation of thought\footnote{The mental trajectory in the phase space dimensions assigned to non-memory tasks can remain chaotic, and could take us to those places (segments of the cylinder, cf.~footnote \ref{fn:cylinder}) where even the motion along the memory dimensions diverges away from the departure projective cycle.}, there is still a good chance that the mind gets entrained onto the transiting paths that take it to the destination projective-cycle. In other words, when we see and remember the cause event, we likely also remember that the consequence event is to follow\footnote{While the consequence event does not have its own direct cue, the fact that mental dynamics would repeatedly get dragged back to the cause event after wandering off, by the cause event's cue, and subsequently frequent also the consequence event via the capacious transiting conduits, offers up a nepotistic indirect control. Thereby the memory of the consequence event is also readily recalled. Heuristically, one could regard the transiting paths as soldering the two projective cycles together, and the cue for the cause event gets recasted into becoming a patchy cue for the larger combined projective cycle.}, thus accomplish a primitive level of prediction, which is the most immediate real world functionality of possessing a sense of causation. Going the other way around, tracing back to the cause after seeing the consequence, appears to be a distinct aptitude only conferred to higher mammals. This is perhaps not surprising, given that the unidirectional transiting paths are likely natural transcriptions of the temporal ordering of the cause and effect events in the real world, thus are procured for free during Hebbian learning. But going the other way requires artificially creating new connecting paths with much effort. 

One may further enlist other types of transit properties to represent negation $\neg$, conjunction $\wedge$, disjunction $\lor$, and etc (the psychological function of binding, namely combining different aspects like color, shape, smell, sound of an object into a wholistic amalgam, may also work in a similar way). Indeed, with improved future understanding of high dimensional hyperchaotic systems, one may even be able to prove that formal mathematical logic exhausts all possible transit varieties available for the biological neural network, and provide a dictionary. One could then argue that mathematics is not only discovered and not invented (there must be a physical neural network substrate substantiating any mathematical contrivance, and we are merely discovering allowed configurations satisfying certain rules), but is in fact accidental, manifesting just a few specific brain functions evolved over eons to help us handle situations arising at the spatiotemporal scale of humans. It is thus not guaranteed to be able to represent and/or process the physics of the extremely small or large scales. For example, the concept of causation relies on the existence of a partial ordering of events, which mirrors the flow of time in the physical world, that in turn depends on the Lorentzian signature of spacetime. However, when investigating particles or cosmology, other signatures may arise, and causation ceases to be a very useful characterization of relationships between events. In particular, there is no reason why formal logic should be self-consistent and omnipotent -- as G\"odel's incompleteness theorem appears to indeed suggest. Encouragingly though, by creating artificial neural networks of more diverse genres, we may be able to expand the logic tool box, and create new paradigms for mathematics and subsequently all quantitative sciences. Practical utility thus behooves us to delve deeper into the realm of higher dimensional hyperchaotic dynamics.

%\newpage
\acknowledgements
This work is supported by the National Natural Science Foundation of China grants 12073005, 12021003, and the Interdiscipline Research Funds of Beijing Normal University. 

%\newpage
 
%\bibliography{../References}
\bibliography{Brain.bbl}

\begin{thebibliography}{57}
\expandafter\ifx\csname natexlab\endcsname\relax\def\natexlab#1{#1}\fi
\expandafter\ifx\csname bibnamefont\endcsname\relax
  \def\bibnamefont#1{#1}\fi
\expandafter\ifx\csname bibfnamefont\endcsname\relax
  \def\bibfnamefont#1{#1}\fi
\expandafter\ifx\csname citenamefont\endcsname\relax
  \def\citenamefont#1{#1}\fi
\expandafter\ifx\csname url\endcsname\relax
  \def\url#1{\texttt{#1}}\fi
\expandafter\ifx\csname urlprefix\endcsname\relax\def\urlprefix{URL }\fi
\providecommand{\bibinfo}[2]{#2}
\providecommand{\eprint}[2][]{\url{#2}}

\bibitem[{\citenamefont{Grebogi and Yorke}(1997)}]{grebogi1997impact}
\bibinfo{author}{\bibfnamefont{C.}~\bibnamefont{Grebogi}} \bibnamefont{and}
  \bibinfo{author}{\bibfnamefont{J.}~\bibnamefont{Yorke}},
  \emph{\bibinfo{title}{The Impact of Chaos on Science and Society}}
  (\bibinfo{publisher}{United Nations University Press}, \bibinfo{year}{1997}),
  ISBN \bibinfo{isbn}{9789280808827},


\bibitem[{\citenamefont{Yao and Freeman}(1990)}]{Yao1990ModelOB}
\bibinfo{author}{\bibfnamefont{Y.}~\bibnamefont{Yao}} \bibnamefont{and}
  \bibinfo{author}{\bibfnamefont{W.~J.} \bibnamefont{Freeman}},
  \bibinfo{journal}{Neural Networks} \textbf{\bibinfo{volume}{3}},
  \bibinfo{pages}{153} (\bibinfo{year}{1990}).

\bibitem[{\citenamefont{Babloyantz and
  Lourenço}(1994)}]{doi:10.1073/pnas.91.19.9027}
\bibinfo{author}{\bibfnamefont{A.}~\bibnamefont{Babloyantz}} \bibnamefont{and}
  \bibinfo{author}{\bibfnamefont{C.}~\bibnamefont{Lourenço}},
  \bibinfo{journal}{Proceedings of the National Academy of Sciences}
  \textbf{\bibinfo{volume}{91}}, \bibinfo{pages}{9027} (\bibinfo{year}{1994}),
  \eprint{https://www.pnas.org/doi/pdf/10.1073/pnas.91.19.9027},

\bibitem[{\citenamefont{Tsuda}(2001)}]{tsuda_2001}
\bibinfo{author}{\bibfnamefont{I.}~\bibnamefont{Tsuda}},
  \bibinfo{journal}{Behavioral and Brain Sciences}
  \textbf{\bibinfo{volume}{24}}, \bibinfo{pages}{793–810}
  (\bibinfo{year}{2001}).

\bibitem[{\citenamefont{Haken and on~Synergetics}(1982)}]{haken1982evolution}
\bibinfo{author}{\bibfnamefont{H.}~\bibnamefont{Haken}} \bibnamefont{and}
  \bibinfo{author}{\bibfnamefont{I.~S.} \bibnamefont{on~Synergetics}},
  \emph{\bibinfo{title}{Evolution of Order and Chaos in Physics, Chemistry, and
  Biology: Proceedings of the International Symposium on Synergetics at Schloss
  Elmau, Bavaria, April 26-May 1, 1982}}, Springer series in synergetics
  (\bibinfo{publisher}{Springer-Verlag}, \bibinfo{year}{1982}), ISBN
  \bibinfo{isbn}{9783540119043},

\bibitem[{\citenamefont{Kitzbichler et~al.}(2009)\citenamefont{Kitzbichler,
  Smith, Christensen, and Bullmore}}]{criticality}
\bibinfo{author}{\bibfnamefont{M.}~\bibnamefont{Kitzbichler}},
  \bibinfo{author}{\bibfnamefont{M.}~\bibnamefont{Smith}},
  \bibinfo{author}{\bibfnamefont{S.}~\bibnamefont{Christensen}},
  \bibnamefont{and} \bibinfo{author}{\bibfnamefont{E.}~\bibnamefont{Bullmore}},
  \bibinfo{journal}{PLoS Comput Biol} \textbf{\bibinfo{volume}{05}},
  \bibinfo{pages}{e1000314} (\bibinfo{year}{2009}).

\bibitem[{\citenamefont{{Zhang} et~al.}(2009)\citenamefont{{Zhang}, {Liu}, and
  {Wang}}}]{ControlChaos}
\bibinfo{author}{\bibfnamefont{H.}~\bibnamefont{{Zhang}}},
  \bibinfo{author}{\bibfnamefont{D.}~\bibnamefont{{Liu}}}, \bibnamefont{and}
  \bibinfo{author}{\bibfnamefont{Z.}~\bibnamefont{{Wang}}},
  \emph{\bibinfo{title}{{Controlling Chaos: Suppression, Synchronization and
  Chaotification}}} (\bibinfo{publisher}{Springer London},
  \bibinfo{year}{2009}), ISBN \bibinfo{isbn}{978-1-4471-2282-1}.

\bibitem[{\citenamefont{Sch{\"o}ll and Schuster}(2008)}]{scholl2008handbook}
\bibinfo{author}{\bibfnamefont{E.}~\bibnamefont{Sch{\"o}ll}} \bibnamefont{and}
  \bibinfo{author}{\bibfnamefont{H.}~\bibnamefont{Schuster}},
  \emph{\bibinfo{title}{Handbook of Chaos Control}}
  (\bibinfo{publisher}{Wiley}, \bibinfo{year}{2008}), ISBN
  \bibinfo{isbn}{9783527622320},

\bibitem[{\citenamefont{Churchland et~al.}(2010)\citenamefont{Churchland, Yu,
  Cunningham, Sugrue, Cohen, Corrado, Newsome, Clark, Hosseini, Scott
  et~al.}}]{churchland2010stimulus}
\bibinfo{author}{\bibfnamefont{M.~M.} \bibnamefont{Churchland}},
  \bibinfo{author}{\bibfnamefont{B.~M.} \bibnamefont{Yu}},
  \bibinfo{author}{\bibfnamefont{J.~P.} \bibnamefont{Cunningham}},
  \bibinfo{author}{\bibfnamefont{L.~P.} \bibnamefont{Sugrue}},
  \bibinfo{author}{\bibfnamefont{M.~R.} \bibnamefont{Cohen}},
  \bibinfo{author}{\bibfnamefont{G.~S.} \bibnamefont{Corrado}},
  \bibinfo{author}{\bibfnamefont{W.~T.} \bibnamefont{Newsome}},
  \bibinfo{author}{\bibfnamefont{A.~M.} \bibnamefont{Clark}},
  \bibinfo{author}{\bibfnamefont{P.}~\bibnamefont{Hosseini}},
  \bibinfo{author}{\bibfnamefont{B.~B.} \bibnamefont{Scott}},
  \bibnamefont{et~al.}, \bibinfo{journal}{Nature neuroscience}
  \textbf{\bibinfo{volume}{13}}, \bibinfo{pages}{369} (\bibinfo{year}{2010}).

\bibitem[{\citenamefont{He}(2013)}]{he2013spontaneous}
\bibinfo{author}{\bibfnamefont{B.~J.} \bibnamefont{He}},
  \bibinfo{journal}{Journal of Neuroscience} \textbf{\bibinfo{volume}{33}},
  \bibinfo{pages}{4672} (\bibinfo{year}{2013}).

\bibitem[{\citenamefont{Schurger et~al.}(2010)\citenamefont{Schurger, Pereira,
  Treisman, and Cohen}}]{schurger2010reproducibility}
\bibinfo{author}{\bibfnamefont{A.}~\bibnamefont{Schurger}},
  \bibinfo{author}{\bibfnamefont{F.}~\bibnamefont{Pereira}},
  \bibinfo{author}{\bibfnamefont{A.}~\bibnamefont{Treisman}}, \bibnamefont{and}
  \bibinfo{author}{\bibfnamefont{J.~D.} \bibnamefont{Cohen}},
  \bibinfo{journal}{Science} \textbf{\bibinfo{volume}{327}},
  \bibinfo{pages}{97} (\bibinfo{year}{2010}).

\bibitem[{\citenamefont{Xue et~al.}(2010)\citenamefont{Xue, Dong, Chen, Lu,
  Mumford, and Poldrack}}]{xue2010greater}
\bibinfo{author}{\bibfnamefont{G.}~\bibnamefont{Xue}},
  \bibinfo{author}{\bibfnamefont{Q.}~\bibnamefont{Dong}},
  \bibinfo{author}{\bibfnamefont{C.}~\bibnamefont{Chen}},
  \bibinfo{author}{\bibfnamefont{Z.}~\bibnamefont{Lu}},
  \bibinfo{author}{\bibfnamefont{J.~A.} \bibnamefont{Mumford}},
  \bibnamefont{and} \bibinfo{author}{\bibfnamefont{R.~A.}
  \bibnamefont{Poldrack}}, \bibinfo{journal}{Science}
  \textbf{\bibinfo{volume}{330}}, \bibinfo{pages}{97} (\bibinfo{year}{2010}).

\bibitem[{\citenamefont{Schurger et~al.}(2015)\citenamefont{Schurger,
  Sarigiannidis, Naccache, Sitt, and Dehaene}}]{doi:10.1073-pnas.1418730112}
\bibinfo{author}{\bibfnamefont{A.}~\bibnamefont{Schurger}},
  \bibinfo{author}{\bibfnamefont{I.}~\bibnamefont{Sarigiannidis}},
  \bibinfo{author}{\bibfnamefont{L.}~\bibnamefont{Naccache}},
  \bibinfo{author}{\bibfnamefont{J.~D.} \bibnamefont{Sitt}}, \bibnamefont{and}
  \bibinfo{author}{\bibfnamefont{S.}~\bibnamefont{Dehaene}},
  \bibinfo{journal}{Proceedings of the National Academy of Sciences}
  \textbf{\bibinfo{volume}{112}}, \bibinfo{pages}{E2083}
  (\bibinfo{year}{2015}),
  \eprint{https://www.pnas.org/doi/pdf/10.1073/pnas.1418730112},

\bibitem[{\citenamefont{SELZ and
  MANDELL}(1991)}]{doi:10.1142/S0218127491000531}
\bibinfo{author}{\bibfnamefont{K.~A.} \bibnamefont{SELZ}} \bibnamefont{and}
  \bibinfo{author}{\bibfnamefont{A.~J.} \bibnamefont{MANDELL}},
  \bibinfo{journal}{International Journal of Bifurcation and Chaos}
  \textbf{\bibinfo{volume}{01}}, \bibinfo{pages}{717} (\bibinfo{year}{1991}).

\bibitem[{\citenamefont{{Zumdieck} et~al.}(2004)\citenamefont{{Zumdieck},
  {Timme}, {Geisel}, and {Wolf}}}]{2004PhRvL..93x4103Z}
\bibinfo{author}{\bibfnamefont{A.}~\bibnamefont{{Zumdieck}}},
  \bibinfo{author}{\bibfnamefont{M.}~\bibnamefont{{Timme}}},
  \bibinfo{author}{\bibfnamefont{T.}~\bibnamefont{{Geisel}}}, \bibnamefont{and}
  \bibinfo{author}{\bibfnamefont{F.}~\bibnamefont{{Wolf}}},
  \bibinfo{journal}{\prl} \textbf{\bibinfo{volume}{93}}, \bibinfo{eid}{244103}
  (\bibinfo{year}{2004}), \eprint{cond-mat/0401038}.

\bibitem[{\citenamefont{Freeman}(1987)}]{freeman1987simulation}
\bibinfo{author}{\bibfnamefont{W.~J.} \bibnamefont{Freeman}},
  \bibinfo{journal}{Biological cybernetics} \textbf{\bibinfo{volume}{56}},
  \bibinfo{pages}{139} (\bibinfo{year}{1987}).

\bibitem[{\citenamefont{Howard et~al.}(2003)\citenamefont{Howard, Rizzuto,
  Caplan, Madsen, Lisman, Aschenbrenner-Scheibe, Schulze-Bonhage, and
  Kahana}}]{howard2003gamma}
\bibinfo{author}{\bibfnamefont{M.~W.} \bibnamefont{Howard}},
  \bibinfo{author}{\bibfnamefont{D.~S.} \bibnamefont{Rizzuto}},
  \bibinfo{author}{\bibfnamefont{J.~B.} \bibnamefont{Caplan}},
  \bibinfo{author}{\bibfnamefont{J.~R.} \bibnamefont{Madsen}},
  \bibinfo{author}{\bibfnamefont{J.}~\bibnamefont{Lisman}},
  \bibinfo{author}{\bibfnamefont{R.}~\bibnamefont{Aschenbrenner-Scheibe}},
  \bibinfo{author}{\bibfnamefont{A.}~\bibnamefont{Schulze-Bonhage}},
  \bibnamefont{and} \bibinfo{author}{\bibfnamefont{M.~J.}
  \bibnamefont{Kahana}}, \bibinfo{journal}{Cerebral cortex}
  \textbf{\bibinfo{volume}{13}}, \bibinfo{pages}{1369} (\bibinfo{year}{2003}).

\bibitem[{\citenamefont{Pesaran et~al.}(2002)\citenamefont{Pesaran, Pezaris,
  Sahani, Mitra, and Andersen}}]{pesaran2002temporal}
\bibinfo{author}{\bibfnamefont{B.}~\bibnamefont{Pesaran}},
  \bibinfo{author}{\bibfnamefont{J.~S.} \bibnamefont{Pezaris}},
  \bibinfo{author}{\bibfnamefont{M.}~\bibnamefont{Sahani}},
  \bibinfo{author}{\bibfnamefont{P.~P.} \bibnamefont{Mitra}}, \bibnamefont{and}
  \bibinfo{author}{\bibfnamefont{R.~A.} \bibnamefont{Andersen}},
  \bibinfo{journal}{Nature neuroscience} \textbf{\bibinfo{volume}{5}},
  \bibinfo{pages}{805} (\bibinfo{year}{2002}).

\bibitem[{\citenamefont{Tallon-Baudry et~al.}(1998)\citenamefont{Tallon-Baudry,
  Bertrand, Peronnet, and Pernier}}]{tallon1998induced}
\bibinfo{author}{\bibfnamefont{C.}~\bibnamefont{Tallon-Baudry}},
  \bibinfo{author}{\bibfnamefont{O.}~\bibnamefont{Bertrand}},
  \bibinfo{author}{\bibfnamefont{F.}~\bibnamefont{Peronnet}}, \bibnamefont{and}
  \bibinfo{author}{\bibfnamefont{J.}~\bibnamefont{Pernier}},
  \bibinfo{journal}{Journal of Neuroscience} \textbf{\bibinfo{volume}{18}},
  \bibinfo{pages}{4244} (\bibinfo{year}{1998}).

\bibitem[{\citenamefont{Wimmer et~al.}(2016)\citenamefont{Wimmer, Ramon,
  Pasternak, and Compte}}]{wimmer2016transitions}
\bibinfo{author}{\bibfnamefont{K.}~\bibnamefont{Wimmer}},
  \bibinfo{author}{\bibfnamefont{M.}~\bibnamefont{Ramon}},
  \bibinfo{author}{\bibfnamefont{T.}~\bibnamefont{Pasternak}},
  \bibnamefont{and} \bibinfo{author}{\bibfnamefont{A.}~\bibnamefont{Compte}},
  \bibinfo{journal}{Journal of Neuroscience} \textbf{\bibinfo{volume}{36}},
  \bibinfo{pages}{489} (\bibinfo{year}{2016}).

\bibitem[{\citenamefont{Spitzer et~al.}(2010)\citenamefont{Spitzer, Wacker, and
  Blankenburg}}]{spitzer2010oscillatory}
\bibinfo{author}{\bibfnamefont{B.}~\bibnamefont{Spitzer}},
  \bibinfo{author}{\bibfnamefont{E.}~\bibnamefont{Wacker}}, \bibnamefont{and}
  \bibinfo{author}{\bibfnamefont{F.}~\bibnamefont{Blankenburg}},
  \bibinfo{journal}{Journal of Neuroscience} \textbf{\bibinfo{volume}{30}},
  \bibinfo{pages}{4496} (\bibinfo{year}{2010}).

\bibitem[{\citenamefont{Jokisch and Jensen}(2007)}]{jokisch2007modulation}
\bibinfo{author}{\bibfnamefont{D.}~\bibnamefont{Jokisch}} \bibnamefont{and}
  \bibinfo{author}{\bibfnamefont{O.}~\bibnamefont{Jensen}},
  \bibinfo{journal}{Journal of Neuroscience} \textbf{\bibinfo{volume}{27}},
  \bibinfo{pages}{3244} (\bibinfo{year}{2007}).

\bibitem[{\citenamefont{Lee et~al.}(2005)\citenamefont{Lee, Simpson,
  Logothetis, and Rainer}}]{lee2005phase}
\bibinfo{author}{\bibfnamefont{H.}~\bibnamefont{Lee}},
  \bibinfo{author}{\bibfnamefont{G.~V.} \bibnamefont{Simpson}},
  \bibinfo{author}{\bibfnamefont{N.~K.} \bibnamefont{Logothetis}},
  \bibnamefont{and} \bibinfo{author}{\bibfnamefont{G.}~\bibnamefont{Rainer}},
  \bibinfo{journal}{Neuron} \textbf{\bibinfo{volume}{45}}, \bibinfo{pages}{147}
  (\bibinfo{year}{2005}).

\bibitem[{\citenamefont{Raghavachari et~al.}(2001)\citenamefont{Raghavachari,
  Kahana, Rizzuto, Caplan, Kirschen, Bourgeois, Madsen, and
  Lisman}}]{raghavachari2001gating}
\bibinfo{author}{\bibfnamefont{S.}~\bibnamefont{Raghavachari}},
  \bibinfo{author}{\bibfnamefont{M.~J.} \bibnamefont{Kahana}},
  \bibinfo{author}{\bibfnamefont{D.~S.} \bibnamefont{Rizzuto}},
  \bibinfo{author}{\bibfnamefont{J.~B.} \bibnamefont{Caplan}},
  \bibinfo{author}{\bibfnamefont{M.~P.} \bibnamefont{Kirschen}},
  \bibinfo{author}{\bibfnamefont{B.}~\bibnamefont{Bourgeois}},
  \bibinfo{author}{\bibfnamefont{J.~R.} \bibnamefont{Madsen}},
  \bibnamefont{and} \bibinfo{author}{\bibfnamefont{J.~E.}
  \bibnamefont{Lisman}}, \bibinfo{journal}{Journal of Neuroscience}
  \textbf{\bibinfo{volume}{21}}, \bibinfo{pages}{3175} (\bibinfo{year}{2001}).

\bibitem[{\citenamefont{Tesche and Karhu}(2000)}]{tesche2000theta}
\bibinfo{author}{\bibfnamefont{C.}~\bibnamefont{Tesche}} \bibnamefont{and}
  \bibinfo{author}{\bibfnamefont{J.}~\bibnamefont{Karhu}},
  \bibinfo{journal}{Proceedings of the National Academy of Sciences}
  \textbf{\bibinfo{volume}{97}}, \bibinfo{pages}{919} (\bibinfo{year}{2000}).

\bibitem[{\citenamefont{Gevins et~al.}(1997)\citenamefont{Gevins, Smith,
  McEvoy, and Yu}}]{gevins1997high}
\bibinfo{author}{\bibfnamefont{A.}~\bibnamefont{Gevins}},
  \bibinfo{author}{\bibfnamefont{M.~E.} \bibnamefont{Smith}},
  \bibinfo{author}{\bibfnamefont{L.}~\bibnamefont{McEvoy}}, \bibnamefont{and}
  \bibinfo{author}{\bibfnamefont{D.}~\bibnamefont{Yu}},
  \bibinfo{journal}{Cerebral cortex (New York, NY: 1991)}
  \textbf{\bibinfo{volume}{7}}, \bibinfo{pages}{374} (\bibinfo{year}{1997}).

\bibitem[{\citenamefont{Hebb}(2005)}]{hebb2005organization}
\bibinfo{author}{\bibfnamefont{D.}~\bibnamefont{Hebb}},
  \emph{\bibinfo{title}{The Organization of Behavior: A Neuropsychological
  Theory}} (\bibinfo{publisher}{Taylor \& Francis}, \bibinfo{year}{2005}), ISBN
  \bibinfo{isbn}{9781135631901},

\bibitem[{\citenamefont{Tsuda}(2009)}]{doi:10.1063/1.3076393}
\bibinfo{author}{\bibfnamefont{I.}~\bibnamefont{Tsuda}},
  \bibinfo{journal}{Chaos: An Interdisciplinary Journal of Nonlinear Science}
  \textbf{\bibinfo{volume}{19}}, \bibinfo{pages}{015113}
  (\bibinfo{year}{2009}), \eprint{https://doi.org/10.1063/1.3076393},

\bibitem[{\citenamefont{Freeman}(2004)}]{Freeman2004SimulationOC}
\bibinfo{author}{\bibfnamefont{W.~J.} \bibnamefont{Freeman}},
  \bibinfo{journal}{Biological Cybernetics} \textbf{\bibinfo{volume}{56}},
  \bibinfo{pages}{139} (\bibinfo{year}{2004}).

\bibitem[{\citenamefont{Skarda and Freeman}(1987)}]{skarda_freeman_1987}
\bibinfo{author}{\bibfnamefont{C.~A.} \bibnamefont{Skarda}} \bibnamefont{and}
  \bibinfo{author}{\bibfnamefont{W.~J.} \bibnamefont{Freeman}},
  \bibinfo{journal}{Behavioral and Brain Sciences}
  \textbf{\bibinfo{volume}{10}}, \bibinfo{pages}{161–173}
  (\bibinfo{year}{1987}).

\bibitem[{\citenamefont{{Mirus} and {Sprott}}(1999)}]{1999PhLA..254..275M}
\bibinfo{author}{\bibfnamefont{K.~A.} \bibnamefont{{Mirus}}} \bibnamefont{and}
  \bibinfo{author}{\bibfnamefont{J.~C.} \bibnamefont{{Sprott}}},
  \bibinfo{journal}{Physics Letters A} \textbf{\bibinfo{volume}{254}},
  \bibinfo{pages}{275} (\bibinfo{year}{1999}).

\bibitem[{\citenamefont{Hopfield}(1982)}]{doi:10.1073-pnas.79.8.2554}
\bibinfo{author}{\bibfnamefont{J.~J.} \bibnamefont{Hopfield}},
  \bibinfo{journal}{Proceedings of the National Academy of Sciences}
  \textbf{\bibinfo{volume}{79}}, \bibinfo{pages}{2554} (\bibinfo{year}{1982}),
  \eprint{https://www.pnas.org/doi/pdf/10.1073/pnas.79.8.2554},

\bibitem[{\citenamefont{Aihara et~al.}(1990)\citenamefont{Aihara, Takabe, and
  Toyoda}}]{AIHARA1990333}
\bibinfo{author}{\bibfnamefont{K.}~\bibnamefont{Aihara}},
  \bibinfo{author}{\bibfnamefont{T.}~\bibnamefont{Takabe}}, \bibnamefont{and}
  \bibinfo{author}{\bibfnamefont{M.}~\bibnamefont{Toyoda}},
  \bibinfo{journal}{Physics Letters A} \textbf{\bibinfo{volume}{144}},
  \bibinfo{pages}{333} (\bibinfo{year}{1990}), ISSN \bibinfo{issn}{0375-9601},

\bibitem[{\citenamefont{{Schuster} and {Stemmler}}(1997)}]{1997PhRvE..56.6410S}
\bibinfo{author}{\bibfnamefont{H.~G.} \bibnamefont{{Schuster}}}
  \bibnamefont{and} \bibinfo{author}{\bibfnamefont{M.~B.}
  \bibnamefont{{Stemmler}}}, \bibinfo{journal}{\pre}
  \textbf{\bibinfo{volume}{56}}, \bibinfo{pages}{6410} (\bibinfo{year}{1997}).

\bibitem[{\citenamefont{CHEN et~al.}(2003)\citenamefont{CHEN, CAO, and
  JIN}}]{doi:10.1142-S0218127403006819}
\bibinfo{author}{\bibfnamefont{H.}~\bibnamefont{CHEN}},
  \bibinfo{author}{\bibfnamefont{Z.}~\bibnamefont{CAO}}, \bibnamefont{and}
  \bibinfo{author}{\bibfnamefont{J.}~\bibnamefont{JIN}},
  \bibinfo{journal}{International Journal of Bifurcation and Chaos}
  \textbf{\bibinfo{volume}{13}}, \bibinfo{pages}{671} (\bibinfo{year}{2003}),
  \eprint{https://doi.org/10.1142/S0218127403006819},

\bibitem[{\citenamefont{He et~al.}(2007)\citenamefont{He, Shrimali, and
  Aihara}}]{he2007partial}
\bibinfo{author}{\bibfnamefont{G.}~\bibnamefont{He}},
  \bibinfo{author}{\bibfnamefont{M.~D.} \bibnamefont{Shrimali}},
  \bibnamefont{and} \bibinfo{author}{\bibfnamefont{K.}~\bibnamefont{Aihara}},
  \bibinfo{journal}{Physics Letters A} \textbf{\bibinfo{volume}{371}},
  \bibinfo{pages}{228} (\bibinfo{year}{2007}).

\bibitem[{\citenamefont{{Amit}}(1987)}]{1987LNP...275..429A}
\bibinfo{author}{\bibfnamefont{D.~J.} \bibnamefont{{Amit}}}, in
  \emph{\bibinfo{booktitle}{Lecture Notes in Physics, Berlin Springer Verlag}},
  edited by \bibinfo{editor}{\bibfnamefont{J.~L.} \bibnamefont{{van Hemmen}}}
  \bibnamefont{and}
  \bibinfo{editor}{\bibfnamefont{I.}~\bibnamefont{{Morgenstern}}}
  (\bibinfo{year}{1987}), vol. \bibinfo{volume}{275}, pp.
  \bibinfo{pages}{429--484}.

\bibitem[{\citenamefont{Ackley et~al.}(1985)\citenamefont{Ackley, Hinton, and
  Sejnowski}}]{Annealing}
\bibinfo{author}{\bibfnamefont{D.~H.} \bibnamefont{Ackley}},
  \bibinfo{author}{\bibfnamefont{G.~E.} \bibnamefont{Hinton}},
  \bibnamefont{and} \bibinfo{author}{\bibfnamefont{T.~J.}
  \bibnamefont{Sejnowski}}, \bibinfo{journal}{Cognitive Science}
  \textbf{\bibinfo{volume}{9}}, \bibinfo{pages}{147} (\bibinfo{year}{1985}).

\bibitem[{\citenamefont{Krotov and Hopfield}(2016)}]{NIPS2016_eaae339c}
\bibinfo{author}{\bibfnamefont{D.}~\bibnamefont{Krotov}} \bibnamefont{and}
  \bibinfo{author}{\bibfnamefont{J.~J.} \bibnamefont{Hopfield}}, in
  \emph{\bibinfo{booktitle}{Advances in Neural Information Processing
  Systems}}, edited by \bibinfo{editor}{\bibfnamefont{D.}~\bibnamefont{Lee}},
  \bibinfo{editor}{\bibfnamefont{M.}~\bibnamefont{Sugiyama}},
  \bibinfo{editor}{\bibfnamefont{U.}~\bibnamefont{Luxburg}},
  \bibinfo{editor}{\bibfnamefont{I.}~\bibnamefont{Guyon}}, \bibnamefont{and}
  \bibinfo{editor}{\bibfnamefont{R.}~\bibnamefont{Garnett}}
  (\bibinfo{publisher}{Curran Associates, Inc.}, \bibinfo{year}{2016}),
  vol.~\bibinfo{volume}{29},

\bibitem[{\citenamefont{Houd{\'e} and
  Tzourio-Mazoyer}(2003)}]{Houd2003NeuralFO}
\bibinfo{author}{\bibfnamefont{O.}~\bibnamefont{Houd{\'e}}} \bibnamefont{and}
  \bibinfo{author}{\bibfnamefont{N.}~\bibnamefont{Tzourio-Mazoyer}},
  \bibinfo{journal}{Nature Reviews Neuroscience} \textbf{\bibinfo{volume}{4}},
  \bibinfo{pages}{507} (\bibinfo{year}{2003}).

\bibitem[{\citenamefont{Gilli}(1993)}]{251826}
\bibinfo{author}{\bibfnamefont{M.}~\bibnamefont{Gilli}}, \bibinfo{journal}{IEEE
  Transactions on Circuits and Systems I: Fundamental Theory and Applications}
  \textbf{\bibinfo{volume}{40}}, \bibinfo{pages}{849} (\bibinfo{year}{1993}).

\bibitem[{\citenamefont{{Kar} et~al.}(2019)\citenamefont{{Kar}, {Kubilius},
  {Schmidt}, {Issa}, and {DiCarlo}}}]{RecurrentCNN}
\bibinfo{author}{\bibfnamefont{K.}~\bibnamefont{{Kar}}},
  \bibinfo{author}{\bibfnamefont{J.}~\bibnamefont{{Kubilius}}},
  \bibinfo{author}{\bibfnamefont{K.}~\bibnamefont{{Schmidt}}},
  \bibinfo{author}{\bibfnamefont{E.}~\bibnamefont{{Issa}}}, \bibnamefont{and}
  \bibinfo{author}{\bibfnamefont{J.}~\bibnamefont{{DiCarlo}}},
  \bibinfo{journal}{Nat Neurosci} \textbf{\bibinfo{volume}{22}},
  \bibinfo{pages}{974} (\bibinfo{year}{2019}).

\bibitem[{\citenamefont{{Musielak} and {Musielak}}(2009)}]{2009IJBC...19.2823M}
\bibinfo{author}{\bibfnamefont{Z.~E.} \bibnamefont{{Musielak}}}
  \bibnamefont{and} \bibinfo{author}{\bibfnamefont{D.~E.}
  \bibnamefont{{Musielak}}}, \bibinfo{journal}{International Journal of
  Bifurcation and Chaos} \textbf{\bibinfo{volume}{19}}, \bibinfo{pages}{2823}
  (\bibinfo{year}{2009}).

\bibitem[{\citenamefont{{Ispolatov} et~al.}(2015)\citenamefont{{Ispolatov},
  {Madhok}, {Allende}, and {Doebeli}}}]{2015NatSR...512506I}
\bibinfo{author}{\bibfnamefont{I.}~\bibnamefont{{Ispolatov}}},
  \bibinfo{author}{\bibfnamefont{V.}~\bibnamefont{{Madhok}}},
  \bibinfo{author}{\bibfnamefont{S.}~\bibnamefont{{Allende}}},
  \bibnamefont{and}
  \bibinfo{author}{\bibfnamefont{M.}~\bibnamefont{{Doebeli}}},
  \bibinfo{journal}{Scientific Reports} \textbf{\bibinfo{volume}{5}},
  \bibinfo{eid}{12506} (\bibinfo{year}{2015}), \eprint{1410.6403}.

\bibitem[{\citenamefont{{Lebowitz} and {Penrose}}(1973)}]{1973PhT....26b..23L}
\bibinfo{author}{\bibfnamefont{J.~L.} \bibnamefont{{Lebowitz}}}
  \bibnamefont{and}
  \bibinfo{author}{\bibfnamefont{O.}~\bibnamefont{{Penrose}}},
  \bibinfo{journal}{Physics Today} \textbf{\bibinfo{volume}{26}},
  \bibinfo{pages}{23} (\bibinfo{year}{1973}).

\bibitem[{\citenamefont{{Kipnis} and {Landim}}(1999)}]{Hydroderivation}
\bibinfo{author}{\bibfnamefont{C.}~\bibnamefont{{Kipnis}}} \bibnamefont{and}
  \bibinfo{author}{\bibfnamefont{C.}~\bibnamefont{{Landim}}},
  \emph{\bibinfo{title}{{Scaling limits of interacting particle systems}}}
  (\bibinfo{publisher}{Springer-Verlag: Berlin, Heidelberg},
  \bibinfo{year}{1999}), ISBN \bibinfo{isbn}{978-3-642-08444-7}.

\bibitem[{\citenamefont{Strogatz}(2007)}]{strogatz2007nonlinear}
\bibinfo{author}{\bibfnamefont{S.}~\bibnamefont{Strogatz}},
  \emph{\bibinfo{title}{Nonlinear Dynamics and Chaos: With Applications to
  Physics, Biology, Chemistry, and Engineering}}, Studies in nonlinearity
  (\bibinfo{publisher}{Levant Books}, \bibinfo{year}{2007}), ISBN
  \bibinfo{isbn}{9788187169857},

\bibitem[{\citenamefont{Wiggins}(1990)}]{wiggins1990introduction}
\bibinfo{author}{\bibfnamefont{S.}~\bibnamefont{Wiggins}},
  \emph{\bibinfo{title}{Introduction to Applied Nonlinear Dynamical Systems and
  Chaos}}, Texts in Mathematics vol. 2 (\bibinfo{publisher}{World Publishing
  Company}, \bibinfo{year}{1990}), ISBN \bibinfo{isbn}{9780387970035},

\bibitem[{\citenamefont{{Zoldi} and {Greenside}}(1998)}]{1998PhRvE..57.2511Z}
\bibinfo{author}{\bibfnamefont{S.~M.} \bibnamefont{{Zoldi}}} \bibnamefont{and}
  \bibinfo{author}{\bibfnamefont{H.~S.} \bibnamefont{{Greenside}}},
  \bibinfo{journal}{\pre} \textbf{\bibinfo{volume}{57}}, \bibinfo{pages}{R2511}
  (\bibinfo{year}{1998}).

\bibitem[{\citenamefont{Devaney and Devaney}(1989)}]{devaney1989introduction}
\bibinfo{author}{\bibfnamefont{R.}~\bibnamefont{Devaney}} \bibnamefont{and}
  \bibinfo{author}{\bibfnamefont{L.}~\bibnamefont{Devaney}},
  \emph{\bibinfo{title}{An Introduction To Chaotic Dynamical Systems, Second
  Edition}}, Addison-Wesley advanced book program (\bibinfo{publisher}{Avalon
  Publishing}, \bibinfo{year}{1989}), ISBN \bibinfo{isbn}{9780201130461},

\bibitem[{\citenamefont{Ivancevic and Ivancevic}(2007)}]{ivancevic2007high}
\bibinfo{author}{\bibfnamefont{V.}~\bibnamefont{Ivancevic}} \bibnamefont{and}
  \bibinfo{author}{\bibfnamefont{T.}~\bibnamefont{Ivancevic}},
  \emph{\bibinfo{title}{High-Dimensional Chaotic and Attractor Systems: A
  Comprehensive Introduction}}, Intelligent Systems, Control and Automation:
  Science and Engineering (\bibinfo{publisher}{Springer Netherlands},
  \bibinfo{year}{2007}), ISBN \bibinfo{isbn}{9781402054563},

\bibitem[{\citenamefont{Cvitanovi\ifmmode~\acute{c}\else
  \'{c}\fi{}}(1988)}]{PhysRevLett.61.2729}
\bibinfo{author}{\bibfnamefont{P.}~\bibnamefont{Cvitanovi\ifmmode~\acute{c}\else
  \'{c}\fi{}}}, \bibinfo{journal}{Phys. Rev. Lett.}
  \textbf{\bibinfo{volume}{61}}, \bibinfo{pages}{2729} (\bibinfo{year}{1988}),

\bibitem[{\citenamefont{{Cvitanovic} et~al.}(1988)\citenamefont{{Cvitanovic},
  {Gunaratne}, and {Procaccia}}}]{1988PhRvA..38.1503C}
\bibinfo{author}{\bibfnamefont{P.}~\bibnamefont{{Cvitanovic}}},
  \bibinfo{author}{\bibfnamefont{G.~H.} \bibnamefont{{Gunaratne}}},
  \bibnamefont{and}
  \bibinfo{author}{\bibfnamefont{I.}~\bibnamefont{{Procaccia}}},
  \bibinfo{journal}{\pra} \textbf{\bibinfo{volume}{38}}, \bibinfo{pages}{1503}
  (\bibinfo{year}{1988}).

\bibitem[{\citenamefont{{Ott} et~al.}(1990)\citenamefont{{Ott}, {Grebogi}, and
  {Yorke}}}]{1990PhRvL..64.1196O}
\bibinfo{author}{\bibfnamefont{E.}~\bibnamefont{{Ott}}},
  \bibinfo{author}{\bibfnamefont{C.}~\bibnamefont{{Grebogi}}},
  \bibnamefont{and} \bibinfo{author}{\bibfnamefont{J.~A.}
  \bibnamefont{{Yorke}}}, \bibinfo{journal}{\prl}
  \textbf{\bibinfo{volume}{64}}, \bibinfo{pages}{1196} (\bibinfo{year}{1990}).

\bibitem[{\citenamefont{Gonz~lez Miranda}(2004)}]{gonz2004synchronization}
\bibinfo{author}{\bibfnamefont{J.}~\bibnamefont{Gonz~lez Miranda}},
  \emph{\bibinfo{title}{Synchronization and Control of Chaos: An Introduction
  for Scientists and Engineers}} (\bibinfo{publisher}{Imperial College Press},
  \bibinfo{year}{2004}), ISBN \bibinfo{isbn}{9781860945229},

\bibitem[{\citenamefont{{Yorke} and {Yorke}}(1979)}]{1979JSP....21..263Y}
\bibinfo{author}{\bibfnamefont{J.~A.} \bibnamefont{{Yorke}}} \bibnamefont{and}
  \bibinfo{author}{\bibfnamefont{E.~D.} \bibnamefont{{Yorke}}},
  \bibinfo{journal}{Journal of Statistical Physics}
  \textbf{\bibinfo{volume}{21}}, \bibinfo{pages}{263} (\bibinfo{year}{1979}).

\bibitem[{\citenamefont{Sparrow}(2012)}]{sparrow2012lorenz}
\bibinfo{author}{\bibfnamefont{C.}~\bibnamefont{Sparrow}},
  \emph{\bibinfo{title}{The Lorenz Equations: Bifurcations, Chaos, and Strange
  Attractors}}, Applied Mathematical Sciences (\bibinfo{publisher}{Springer New
  York}, \bibinfo{year}{2012}), ISBN \bibinfo{isbn}{9781461257677},

\end{thebibliography}

\appendix
\section{Prevalence of chaos \label{sec:prevalence}} 
Seen as a dynamical system, the phase space of the brain has a dimension equaling the number of neurons\footnote{If we only count the firing of synapses in a binary fashion, then each neuron only has two discrete states. So we would have a continuous in time system with a discrete phase space. However, the amplitude of the firing also matters, so we instead end up with a continuous phase space of voltages, whose dimension is as large as the number of neurons. This continuity requirement is not superficial, since jumpy dynamics are not constricted by non-intersecting requirements on trajectories, allowing even wilder behaviour, much like how discrete systems can be chaotic even in 1-D. In particular, such continuity (bounded derivatives) helps satisfy a minimal requirement called the Lipschitz condition, which ensures the existence and uniqueness of the dynamical evolution.}\footnote{The brain is not homogeneous. Different regions of the brain carry out different tasks, and in phase space, each region corresponds to a set of dimensions, so the inhomogeneity manifests as anisotropy in phase space, allowing the strange attractors to only take up certain orientations.} (see e.g., \cite{251826}), which is in the tens of billions. Even in the lower dimensional cases, we have already seen from existing literature, that the behaviour of a dynamical system is highly variable and sensitively dependent on a vast array of factors\footnote{Note that the phase space construction does not take into account the connections between neurons, which is imposed by the dynamics instead, via e.g., coupling strength coefficients. In addition to feedback loops (cf., recurrent networks \cite{RecurrentCNN}) reflected in these coefficients, there would generally also be time delays, introducing path dependency and thus further enhance the apparent randomness (i.e., the chaotic nature) of the neural network. See e.g., Eq.~7.39 of \cite{ControlChaos} for the dynamic equations of an example artificial network.} such as number and power of the nonlinear terms, the symmetries or parameter settings. For the even higher dimensional case, the chance that we can conjure up a vastly simplified toy model that happens to provide a good description of brain functions is null. Therefore, we shy away from quantitative modeling, since it is likely more misleading than illuminating, and instead concentrate on generic qualitative discussions that likely apply to most, if not all, sufficiently complex systems. To truly understand the human mental functions at a quantitative level, neurobiology will have to be examined in exquisite detail. Alternatively, if one is interested in creating artificial general intelligence that is superior to humans, it may be more profitable to forget about evolutionary flukes and instead tailor-design its neural networks according to mathematical understanding of higher dimensional systems. 

Unfortunately though, studies on very high dimensional dynamical systems are lacking\footnote{Studies on chaos rely heavily on numerics, and the multibillion dimensional brain is not easy to simulate.} \cite{2009IJBC...19.2823M}, especially so on the hyperchaotic\footnote{More than one positive Lyapunov exponents.} ones, and on those involving attractors or cycles that are projections of trajectories in an even higher dimensional phase space\footnote{The projected orbits can intersect, invaliding among others, the topological tools.}. Nevertheless, it is not difficult to argue that chaos is generic. Indeed, it is shown by Ref.~\cite{2015NatSR...512506I} that in dimensions higher than about fifty, systems are chaotic with probability essentially unity. One may also note that chaos, being implied by ergodicity and the even stronger mixing properties \cite{1973PhT....26b..23L}, is necessary for statistical mechanics to be valid for large many-particle systems (see e.g., \cite{1973PhT....26b..23L}), thus the fact that statistical treatments appear to work for physics experiments\footnote{Neuroscience is a little more complicated. Typically, when the number of degrees of freedom goes very large, we would resort to statistical treatments. From the thermodynamic point of view though, a living brain capable of hosting complicated thoughts would require a very structured and fragmented phase space, rather than being in thermal equilibrium -- which is just one giant indistinguishable monoculture. In other words, it is desirable for the brain to be extremely far away from thermo-equilibrium, and with no quick way to get there before the end of life. Even worse, the fact that the deterministic chaos does not resemble randomness in the case of the brain likely precludes even non-equilibrium thermodynamic treatments, which often assume stochasticity (see e.g., \cite{Hydroderivation}). In any case, while some degree of averaging may be relevant when deciphering the brain's power to handle incomplete or fuzzy information, the type of extreme coarse-graining that would reduce the effective degrees of freedom within the brain from billions to just a few, so as to be treatable in the usual thermodynamic fashion, does not seem to leave sufficiently numerous details to account for psychological intrigues.} lends support for the generality of chaos in high dimensional phase spaces. Furthermore, strange attractors likely readily exist\footnote{We disregard Hamiltonian chaos since the brain is a highly dissipative system that generates heat, so the comoving phase space volume generally contracts (i.e., the system can become confined onto lower Hausdorff dimensional objects like the strange attractors). For the same reason, quasi-periodic orbits wrapped around high-dimensional homotopically-nontrivial submanifolds would be difficult to sustain (as the volume of the manifold should generally drop with increasing time due to dissipation, then the orbits wrapped around it experience a secular change and are not really ``quasi-periodic'' anymore, see e.g., Example 9.2.1 in \cite{strogatz2007nonlinear}), but those wrapped around low (much lower than the phase space, so it being invariant does not prevent an enclosing volume at the dimension of the phase space from shrinking) dimensional submanifolds might exist (see e.g., \cite{2009IJBC...19.2823M} for summary of some examples). For brevity, we don't explicitly distinguish such lower dimensional (small number of incommensurate frequencies) quasi-periodic orbits, and refer to them as being ``periodic'', since they are also confined to thin strata of the phase space and thus can be very specific in what information they represent, just like the exact periodic orbits.}, since the phase trajectories are alway 1-D curves, which in higher dimensions would enjoy more headroom to wind around each other without intersecting, even when confined to a limited volume. 

\section{Suppressing chaos \label{sec:control}}
Control theory starts by noting that the strange attractors in a chaotic system typically contain dense collections of unstable cycles (periodic orbits) \cite{wiggins1990introduction,1998PhRvE..57.2511Z}. In fact, it is one of the conditions for chaos in the definition of Devaney \cite{devaney1989introduction}. Intuitively, the strange attractors are trapping just like cycles are, yet most trajectories in it never close and are instead transitive/ergodic, but these can nevertheless be regarded as belonging to the infinite period limit for closed cycles. The strange attractor will be filled mostly by such orbits, with some short period orbits sprinkled here and there. As we include longer and longer periods into our consideration though, those interstitial periodic cycles become denser and denser (number of orbits grows exponentially with period, see the Smale horseshoe discussion in e.g., \cite{ivancevic2007high}), thus more accurately sample and consequently depict the outline of the strange attractor, much like a pointillism painting (Ref.~\cite{PhysRevLett.61.2729} offers actionable procedures for corporealizing this finite cycle approximation). If we include all finite period cycles, then we acquire a dense subset (a skeleton, see e.g., \cite{1988PhRvA..38.1503C}) that almost accurately reproduce the entire strange attractor.  

Subsequently, if we introduce controls into the system in such a way that some of the shorter period cycles get stabilized, then chaos can be suppressed \cite{1990PhRvL..64.1196O,gonz2004synchronization}. For an extreme example that illustrates intuition, let each control be as strong as a constraint that confines the system onto a lower dimensional submanifold in the phase space, much like what energy conservation does, then if we reduce the dimensionality all the way to two by laying down sufficient number of controls, the Poincar\'e-Bendixson theorem would assert that there cannot be chaos and cycles become likely. Alternatively, we could have something like the transition between a strange attractor and either the sleuth of saddle-cycles (the transient chaotic regime) in the Lorenz system \cite{1979JSP....21..263Y}, or the simpler global cycle at the other side of chaos in the control parameter space \cite{sparrow2012lorenz}. 

A character altering scenario such as this turns out to be relatively easy to realize, because first of all, whatever the specific trajectory of a chaotic system, it can always get within intimate neighborhoods of any target cycle, thanks to ergodicity \cite{scholl2008handbook}. Therefore, once the trajectory gets close to a target orbit, all we need to do for stabilization is to fine-tune some accessible control parameters and/or the dynamical variables, in such a way as to curb the tendency to diverge away from it again along some directions\footnote{The periodic orbits are unstable if we do nothing. In that case the diverging away of nearby trajectories is described by Smale horseshoe maps on disks transverse to the periodic orbit. Only a fractal set of points stay close over time.}. Such a task is easier if we have at hand a reference system to compare with, that is similar to the desired period orbit, so we have a measure of the digression from said cycle, and then some sort of negative feedback can be applied to drive down the deviation. 

As an example, we can take a look at the Adaptive-Reference Adaptive Control method \cite{ControlChaos}. Suppose we have a chaotic dynamical system  
\bea
\dot{\bf x} = \mathcal{F}({\bf x},t,{\boldsymbol \mu})\,,
\eea 
where ${\bf x} \in \mathbb{R}^n$ is the state vector, while $\boldsymbol \mu \in \mathbb{R}^m$ is some control parameter set, and we would like to change the dynamics by moving $\boldsymbol\mu$ to some $\boldsymbol\mu_g$ such that some originally unstable cycle becomes stabilized. Typically, one can not explicitly measure the present state of $\boldsymbol\mu$ (these are implicit control parameters) or indeed know exactly what $\boldsymbol\mu_g$ should be, thus has no means to just drive changes in the parameters directly. Nevertheless, we have access to the explicit state variable ${\bf x}$, as well as some reference ${\bf y}$ orbit that satisfies the desired dynamics corresponding to $\boldsymbol\mu_g$ , i.e., we have
\bea
\dot{{\bf y}} = \mathcal{F}_c({\bf y},t,\boldsymbol\mu_g)\,,
\eea
where $\mathcal{F}_c =\mathcal{F}$ when the reference is of good quality. Consequently, we could compare the two and use the error function ${\bf e} \equiv {\bf x}-{\bf y}$ as a gauge of how far we are from the desired controlled target system, and design a feedback mechanism to drive $\boldsymbol\mu$ in a direction that suppresses ${\bf e}$. In the case when ${\bf y}$ is fairly coherent, e.g., when it is an equilibrium point or some short period cycle, ${\bf e}$ evanesces away quite consistently and this method can be applied as is. When ${\bf y}$ itself is chaotic or almost so (very long period cycles) however, the norm $\|{\bf e}\|$ (usually some $L_{\infty}$ norm) can swing wildly and the control typically becomes less efficient. 

\end{document}